\documentclass[12pt,a4paper,reqno
]{amsart}
\usepackage{graphicx}
\usepackage{amsmath,amssymb,amsthm,array}
\usepackage{amsfonts, amsmath, amsthm, amssymb}
\newcommand{\sign}{\mathop{\mathrm{sign}}\nolimits}%

\newtheorem{assumption}{Assumption}

\newcommand{\LambertW}{\mathop{\mathrm{LambertW}}\nolimits}
\newcommand{\sech}{\mathop{\mathrm{sech}}\nolimits}
\newcommand{\arcsech}{\mathop{\mathrm{arcsech}}\nolimits}

     \title[ Shock waves of spherical and cylindrical KdV-B] {Shock waves of spherical/cylindrical KdV-B:
     Asymptotic, stability, superposition. }%
 \author{Alexey Samokhin}\vspace{6pt}
\address{Institute of Control Sciences of Russian
Academy of Sciences
65 Profsoyuznaya street, Moscow 117997, Russia}\vspace{6pt}
\email{ samohinalexey@gmail.com}\vspace{6pt}
\begin{document}
\maketitle

\begin{abstract}
 Spherical and cylindrical KdV-B equations  have few known exact solutions, yet these solutions are hard to be interpreted physically. But these equations do have a    family of diverging shock waves. Their properties such as
 asymptotic modes, stability, rules of  their interactions/superposition are the subject of this paper. It gives a detailed asymptotic description of the one-parameter families of shock wave solutions and proves their stability using a conservation law. Based on these results, effective rules of superposition are obtained. Moreover these rules are applicable to a  wide class of shock waves,  in particular discontinuous.
Typical examples are illustrated by   graphs.
\end{abstract}



\maketitle

\section*{Introduction}

The spherical and cylindrical KdV-B equations  are of the form

\begin{equation}\label{1}
u_t  =-2uu_x -n\frac{u}{t} + \varepsilon u_{xx}+\lambda u_{xxx}.
\end{equation}

Here $\lambda>0$, $\varepsilon>0$ are constants connected to dispersion
and dissipation; $n=0,\;1,\;\frac{1}{2}$ for the flat, spherical and cylindrical species respectively.

The combination of dissipation and dispersion in \eqref{1} produce families of continuous monotonic or oscillatory shock waves as it is shown below.

       Throughout this paper  shock waves diverging from the center
       (continuous as well as "real" shocks, i.e., waves  with discontinuities) are assumed to have the  following boundary condition
       \[
       u|_{x=+\infty}=0,\quad u|_{x=+0}=H.
       \]
      . Such waves and their properties are the subject of this paper.

There is a considerable number of publication in the field.  Here  we cite the most relevant ones.
  In \cite{1} the interaction of the BBM equation's shock waves is  deduced using the momentum conservation law decay. In \cite{2}
the interaction of the BBM  solitary waves is  described also using conservation laws.  In \cite{3} a superposition of the flat KdV-B travelling waves is discussed.
 The review \cite{04} deals with modelling Rieman-type shocks for
$u_t=f(u)_x$ type equations,
with modified KdV (mKdV) and flat KdV-B as a main examples; it also contains a vast bibliography.

Equations \eqref{1} have numerous applications, but few their
exact/explicit solutions are known.

For instance, the non-dissipative (e.g., for $\varepsilon=0$) equations \eqref{1} have solitary wave solutions.
Exact solution for such a  cylindrical KdV-B equation
can be obtained by making a simple substitution
of the unknown function and the independent variable, see \cite{4}.
The result is the autonomous KdV-B equation of the standard
form

\begin{equation}\label{KdV}
u_t=-2uu_x\lambda u_{xxx}.
\end{equation}
Some of its exact solutions (e.g. solitons) are known. Returning to the original variables,
one can obtain an exact solution to the original equation. This was done \cite{5}.
Yet, the solution \eqref{1} obtained in this way, have no  physical interpretation, since its
velocity profile increases linearly as
$x\rightarrow +\infty$, and for some $x$ the velocity can even be
negative.

In \cite{6}, it is argued that it is impossible to derive another exact solution
to equation \eqref{1} that could have physical meaning, and therefore, some numerical models must be used.
Numerical studies of equation  \eqref{1} were conducted in
 in \cite{7} and a simple approximation formula was proposed that describes
the behavior of a cylindrical soliton.

For dissipation-less ($\epsilon=0$) symmetry invariant solution must be of the form
\[
u(x,t)=x^{-1}f(\frac{t}{x^2}),
\]
see \cite{9}; but the corresponding ordinary differential equation seems too complex.

In \cite {10} the uniform stability of oscillatory shocks for flat KdV-Burgers equation ($n=0$ case) is analyzed  by a complex approach that is radically different from ours.

This paper content is as follows.

Section \ref{types} contains an effective asymptotic description of the spherical and cylindrical KdV-B
continuous shock waves  in terms of their boundary condition.

In section \ref{stable} the stability of
continuous  shocks is proven
using the selective decay rate (see e.g, \cite{6}), of the KdV \emph{momentum} conservation law.

   In section \ref{super}  simple  rules are given for
superposition of  travelling shocks, both  continuous and discontinuous.

The section \ref{discus} states the main results and contain some ideas for further research.

\section{Shock wave solutions\label{types}}

\subsection{Flat KdV-B}

The travelling wave solutions  are of the form

\begin{equation}\label{2}
u = u(x - V t).
\end{equation}
The graph of such  solution stay unaltered while travelling along the $x$-axis with velocity $V$.

In the flat case   $n=0$, that is for the well studied Kdv-B, equation
\begin{equation}\label{3}
u_t  =-2uu_x + \varepsilon u_{xx}+\lambda u_{xxx}.
\end{equation}
 a family of exact travelling wave solutions is known:

\begin{eqnarray}\label{4}
  u(x,t)=u(x-Vt)=\frac{V}{2}+\nonumber\\
  \frac{3\varepsilon^2}{50\lambda}\left[
  \tan^2\left(\frac{3\varepsilon}{10\lambda}(x-Vt+s)\right)\pm
  2\tan\left(\frac{3\varepsilon}{10\lambda}(x-Vt+s)\right)-1\right]
,
  \end{eqnarray}
 $V$ being    the parameter and $s$ --- an arbitrary translation along the $x$ axis.

In the fixed medium $\{\varepsilon, \lambda\}$ it describes a one-parameter family denoted below as
$F_{V,s}(x,t)$ where $V \in \mathbb{R}$;  $s$ is  arbitrary; it is usually omitted for brevity.

Generally, substitution of \eqref{2} into \eqref{3} results in ODE
  \begin{equation}\label{5}
-V u_x + 2uu_x  \lambda V u_{xxx} -\varepsilon u_{xx} = 0.
\end{equation}

 The expression \eqref{4} can be obtained by twice lowering the order of \eqref{5}, see \cite{3}.
The integration of \eqref{4} leads to

  \begin{equation}\label{05}
-V u +u^2-\lambda  u_{xx}-\varepsilon u_x=C
\end{equation}

 Instead of $V$ and $C$ in \eqref{05} , it is convenient to chose
 the boundary conditions  $u(x)|_{x=\pm \infty}$ as the other two parameters.

\begin{assumption}
For any  KdV-B travelling shock wave of the following property is assumed
 \begin{eqnarray}
\nonumber(\exists\; u(x)|_{x=-\infty}=H)                                     \; &\wedge &\; (\exists\; u(x)|_{x=+\infty}=h)\\
\bigwedge (\forall n>0: \; \frac{d^nu(x)}{dx^n}|_{x=\pm \infty}=0).&&
  \end{eqnarray}
\end{assumption}

This is analogical to a property of rapidly decreasing functions
\[\forall n\in\mathbb{N}: \; \frac{d^nu(x)}{dx^n}|_{x=\pm \infty}=0,
\]
common in mathematical  physics when describing a
localized phenomena (e.g. solitons).

a (e.g. solitons).

Applying the above assumption to \eqref{5} we get

\begin{equation}\label{6}
  \left\{
    \begin{array}{l}
      -Vh+h^2=C,   \\
      -VH+H^2=C.
    \end{array}
  \right.
\end{equation}

It follows that

\begin{equation*}
V=H+h, C=-Hh
\end{equation*}

Hence a family of the flat KdV-B shock  (denoted below $ T_{H,h}(X-Vt+s)$) is  2-dimensional.
In particular, for the explicit TWS \eqref{4}
\[
S_{V,s}=T_{\frac{V}{2}\pm\frac{3\varepsilon^2}{25\lambda}, \frac{V}{2}\mp \frac{3\varepsilon^2}{50\lambda}}.
\]
So
\begin{equation}\label{7}
\mbox{If }h=0\mbox{ then }V=H,\; C=0
\end{equation}
 and namely these solutions are important in this study: the particular case $h=0$  is connected with cylindrical and spherical shocks.
The notation
$F_{V,s}(x-Vt+s)=T_{V,0}(x,t)$ is used below.
Note that in this case  the velocity of the wave is numerically equal to its  height.

For the exact solution $S_{V,s}$ of \eqref{4} $V=\frac{6\varepsilon^2}{25\lambda}$ holds.

To describe the structure of this $\{(H,0)=(V,0)\}$ family consider  the dynamical system associated with \eqref{5}:

\begin{equation*}
  \begin{array}{rcl}
    u'&= & p \\
    \lambda p'&=& Vu -u^2- \varepsilon p.
  \end{array}
\end{equation*}

The fixed points satisfy  conditions $u'=0,; p'=0$, so by \eqref{7}
$Vu -u^2+C=0$ and $C=0$ It follows
that the fixed points are, naturally, $(V,0)$ and $(0,0)$.

Next step is to find the types of these  points. Solve the characteristic equation

\begin{equation*}
  \det
  \begin{pmatrix}
           \frac{\partial u'}{\partial u}-k & \frac{\partial u'}{\partial p} \\
           \frac{\partial p'}{\partial u} & \frac{\partial p'}{\partial p}-k \\
         \end{pmatrix}=\det
  \begin{pmatrix}-k&1\\
  \frac{V-2u}{\lambda }& \frac{\varepsilon}{\lambda }-k
         \end{pmatrix}=0,
\end{equation*}

that is

\begin{equation*}
  k^2-\frac{\varepsilon}{\lambda }k-\frac{(V-2u)}{\lambda }=0.
\end{equation*}

The roots are

  \begin{equation*}
  k_\pm =\frac{\varepsilon}{2\lambda } \pm
  \sqrt{\left( \frac{\varepsilon}{2\lambda }\right)^2+
\frac{(V-2u)}{\lambda}}\quad u=0;V.
\end{equation*}

The real and complex value roots are separated by  the sign of the discriminant

  \begin{equation*}
\left( \frac{\varepsilon}{2\lambda }\right)^2+
\frac{(V-2u)}{\lambda }=\frac{1}{\lambda}
\left( \frac{\varepsilon^2}{4\lambda}\pm V\right)
\end{equation*}

(the choice between $"+"$ and $"-"$ here depends on the choice of a fixed point).

For $V$ big  enough the fixed point $(0,0)$ has complex roots
\[
k_\pm =\frac{\varepsilon}{2\lambda } \pm
  \imath\sqrt{\frac{V}{\lambda }-\left( \frac{\varepsilon}{2\lambda }\right)^2
}
\]
 this point corresponds to a focus on the phase portrait of the dynamical system. Related solution of the flat KdV-B is of oscillatory type.

  Another point then has $k_\pm$ real a and of different signs; it is a saddle.  both fixed points have real roots
Uf
The the corresponding solution is of a monotonic type; it connects fixed points and defines a separatrix on the directional field.

For  smaller $V$ all roots are real and the fixed points are  saddle and unstable node. Then  the separatrix connecting them corresponds to a  monotonic solution.

Hence the criterion "monotonic versus oscillatory" solution  is

  \begin{equation*}
  \sign\left[V- \frac{\varepsilon^2}{4\lambda}\right].
 \end{equation*}

Recall that for exact solution \eqref{4}, $V=\frac{6\varepsilon^2}{25\lambda}$, so this wave is monotonic.

\subsection{Asymptotic for cylindrical and spherical shocks\label{asymp}}

There were few attempts, so far unsuccessful, to obtain physically interpretable exact/analitical solution for these equations.

Below we describe the KdV-B shocks, both
spherical and cylindrical, by means of asymptotics (cf. \cite{9} --- analogous approach, but containing important mistakes).

To be precise, the following properties on solutions are assumed:
\begin{itemize}
  \item solutions are define on $(+0,+\infty)$;
  \item $\exists\; u(x)|_{x=+0}=H \; \wedge \; \exists\; u(x)|_{x=+\infty}=h)$
  \item $\forall n \in \mathbb{N}: \; \frac{d^nu(x)}{dx^n}|_{x=
+\infty}=0).$
\end{itemize}

In fact, $u(x)|_{x=+\infty}=h=0$. Indeed,  all summands  of \eqref{1} at
$x=+\infty$ are zero, except for $u_t=-n\frac{u}{t}$. So $u_{+\infty}t^n=ht^n=const, \; n\neq 0$ which is possible only for $h=0$.

\begin{figure}[h]
\begin{minipage}{13pc}
\includegraphics[width=13pc]{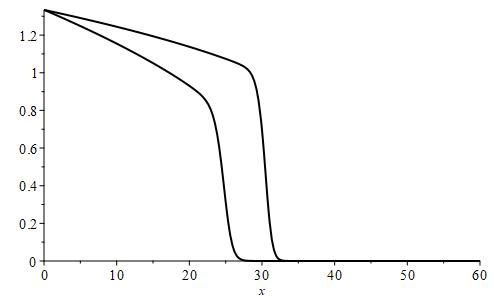}
\end{minipage}\hspace{2pc}%
\begin{minipage}{13pc}
\includegraphics[width=13pc]{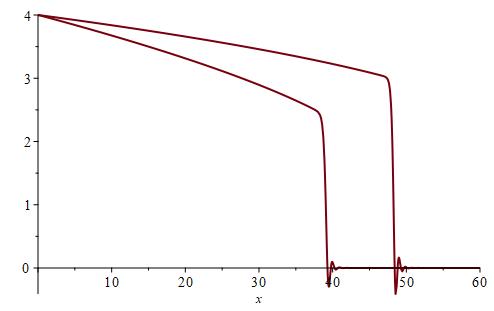}
\end{minipage} 
\caption{\small \textbf{Left }Monotonic $V=\frac{4}{3}$ \textbf{Right} oscillatory shocks. The lower line is spherical, upper - cylindrical on both graphs.}
\label{compare}
\end{figure}

Numerical experiments stably demonstrate two  patterns,monotonic and oscillatory, see Figure \ref{compare} in the case of
; $V=3$ $\{\varepsilon=0.5,\lambda=0.05\}$

The graphs are as if compiled of two parts: at vicinity of
zero it is a convex monotonically decreasing line; the rest part (shock front and over) seems identical with to a graph of flat KdV shock.

\begin{figure}[h]
\begin{minipage}{13pc}
\includegraphics[width=13pc]{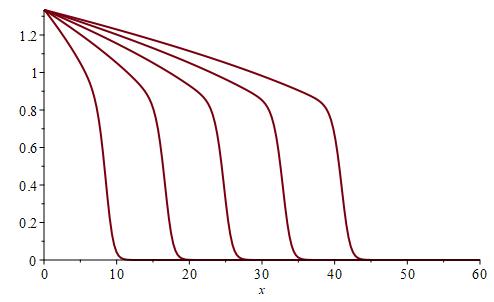}
\end{minipage}\hspace{2pc}%
\begin{minipage}{13pc}
\includegraphics[width=13pc]{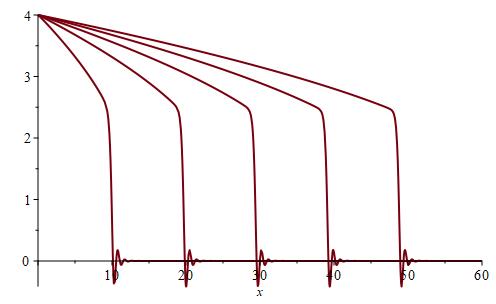}
\end{minipage}2
\caption{\small \textbf{Left }Monotonic spherical shock $\{\varepsilon=0.5,
\lambda=0.05\},$ $t=10,20,30,40,50$,$V=1$; \textbf{Right} oscillatory
$\{\varepsilon=0.,
\lambda=0.05\},$ $t=4,8,12,16$, $V=3$.}
\label{dynamics}
\end{figure}

If seen in dynamics,
 solution's graph animation one can clearly see (Figure \ref{dynamics}, right) that the convex monotonic part before its head shock develops as a nomothetic transformation with a constant velocity $V$.

So for the first part we seek solutions of the form $u(x,t)=y(\frac{x}{Vt})$. Substituting it into equation \eqref{1} we get the equation

\begin{equation*}
 -y'\frac{x}{Vt^2}+\frac{ny}{t}=\frac{2yy'}{Vt}+\frac{\varepsilon^2 y''}{(Vt)^2}+\frac{\lambda y'''}{(Vt)^3},
\end{equation*} 
or
\begin{equation*}
 -\xi y'+ny=-\frac{2yy'}{V}+\frac{\varepsilon^2 y''}{V^2t}+\frac{\lambda y'''}{V^3t^2}.
\end{equation*}
for $y=y(\xi)$.
As $ \xi$ is the independent  variable of ordinary differential equation while $t$ is an arbitrary parameter,  two last summands are omitted:
\[
 -\xi y'+ny=\frac{2yy'}{V}.
\]
In order to obtain  appropriate solution  of the resulting first-order equation an initial condition is needed.

Let $V$ is the velocity of the shock propagation in the medium . Since at the head  shock $x=Vt$ and $u=V$,  so the condition is $ y(V)=V$.
 
It follows then that
for  the cylindrical waves the monotonically decreasing convex part is given by 
\[u=\frac{1}{3}\left(2V+V\sqrt{4 -\frac{3x}{Vt}}\right);
\]
for spherical waves
\[
u=V\sqrt{e}\exp\left(\LambertW\left(-\frac{x}{2 V t\sqrt{e}}\right)\right).
\]

Note that, by the above formulas,
\[
u_{cyl}(+0)=H=\frac{4}{3}V,\; u_{sph}(+0)=H=\sqrt{e}\cdot V,
\] 
so
\begin{equation*}
V =\frac{3}{4}H\mbox{for cylindrical shock
for spherical }V=\frac{H}{\sqrt{e}}.
\end{equation*}

These formulas show dependence of velocity on the amplitude at the start of a wave.

For $t$ big enough the terms $n\frac{u}{t}$ in \eqref{1} may be omitted. Thus we come to the flat KdV-B and therefore the rest  part of solutions
(its shock front and over) is a  part of flat KdV-B  travelling wave.
Hence the first idea of an approximation is as follows:

\begin{equation*}
  u_{cyl}(x,t)=\left\{
     \begin{array}{ll}
      \frac{1}{3}\left(2V+V\sqrt{4 -\frac{3x}{Vt}}\right) , & x\leq Vt; \\
       F_{V,s}(x-Vt+s)   ,  & x> Vt.
     \end{array}
   \right.
\end{equation*}
 and analogously for spherical shocks.

Yet  these formulas need  corrections.

\begin{figure}[h]
\begin{minipage}{13pc}
\includegraphics[width=13pc]{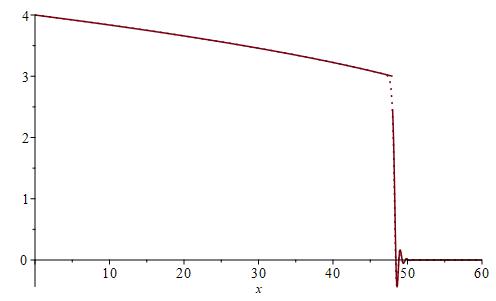}
\end{minipage}\hspace{2pc}%
\begin{minipage}{13pc}
\includegraphics[width=13pc]{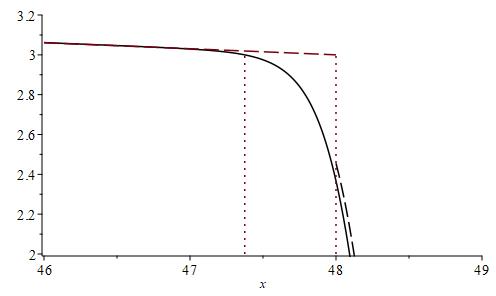}
\end{minipage}
\caption{\small \textbf{Left } Truncation needed, $r_*=s_*=0$; dot line is a missing part of cylindrical shock.
\textbf{Right} Enlarged part at a vicinity of the conjugation point $(Vt,V)$; dash lines --- approximation.}\label{initial}
\end{figure}

If the zero initial conditions are used then:
\begin{itemize}
\item the shock front of a cylindrical wave starts before the point $x=Vt$;
 \item at a point $x=Vt$ the height  $F_{V,0}(x-Vt) =F_{V,0}(0) $ corresponds to a middle of the shock front, not to its highest start and it leads to a  break, see Figure \ref{initial}.
\end{itemize}
So the first correction  is to truncate the region of the formula
$ \frac{1}{3}\left(2V+V\sqrt{4 -\frac{3x}{Vt}}\right)$ action  by some $r_*$, while simultaneously widened for $F_V$.

 A good estimate of $r_*$ may  be obtained using a travelling
shock of the Burgers equation, $u_t=-2uu_x+\varepsilon u_{xx}$: its shocks are very similar to that of the flat KdV-B,  have velocity and height $V$. The  analytical form of a Burgers shock is
\[
w(x,t)=\frac{V}{2}\left[1-\tanh\left(\frac{V}{2\varepsilon}(x-Vt+s)
\right)\right].
\]
and
\[
w_x|_{t=s=0}=\frac{V^2}{4\varepsilon}\sech^2\left(\frac{Vx}
{2\varepsilon}\right).
\]
Now $r_*$ is the root of the equation $w_x=\theta$ where $\theta$ is
sufficiently small (in example below, $\theta=0.01$). Then

\begin{equation}\label{r*}
r_*\approx \frac{2\varepsilon}{V}\arcsech\left(\frac{2\varepsilon}{V}
\sqrt{\varepsilon\theta}\right);
\end{equation}

The  cause of the second correction is following: in  numerical experiments, the shock front of the flat KdV-B wave may occur slightly shifted with respect to the cylindrical  wave. Since the latter cannot be shifted, a very small  shift $s_*$ (about $10^{-3}$)
may be introduced  into a flat shock (recall that it is translation invariant). The origin of this shift may be  a numerical artefact. The same correction may be applied  in the spherical case.

On Figure \ref{initial} (right) the truncation  is shown by dot lines.

The following Figures \ref{initial} and \ref{final} are results of a cylindrical wave simulation in the media $\{\varepsilon=0.2,\lambda=0.05\},$ for $t=16, V=3$.

\begin{figure}[h]
\begin{minipage}{13pc}
\includegraphics[width=13pc]{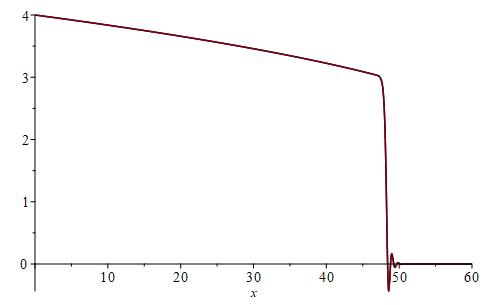}
\end{minipage}\hspace{2pc}%
\begin{minipage}{13pc}
\includegraphics[width=13pc]{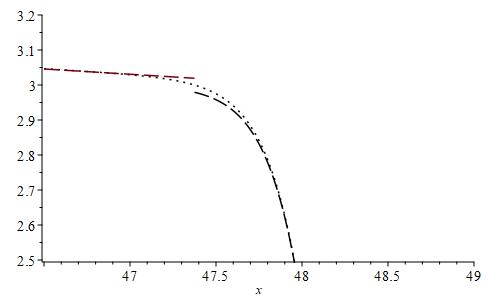}
\end{minipage}
\caption{\small \textbf{Left } Proper
conjugation, $r_*=0.63,s_*$; \textbf{{Right}} Enlarged part at a vicinity of the conjugation point $(Vt,V)$; dash lines - approximation, dots - cylindrical KdV-B.}\label{final}
\end{figure}

 The final approximation now reads
\begin{equation}\label{appC}
  u_{cyl}(x,t)=\left\{
     \begin{array}{ll}
      \frac{1}{3}\left(2V+V\sqrt{4 -\frac{3x}{Vt}}\right), &
x< Vt-r_*; \\
       F_{V,s}(x-Vt+s_*)   ,  & x\geq Vt-r_*.
     \end{array}
   \right.
\end{equation}

\begin{equation}\label{appS}
  u_{sph}(x,t)=\left\{
     \begin{array}{ll}
      u_3=V\sqrt{e}\exp\left(\LambertW\left(-\frac{x}{2 V t\sqrt{e}}\right)\right)& x< Vt-r_*.
 \\
    F_{V,s}(x-Vt+s_*)   , & x\geq Vt-r_*.
     \end{array}
   \right.
\end{equation}
In formulas \eqref{appC} and \eqref{appS} $F_{V,s}$ is constructed with zero initial condition.

After the shift and truncation corrections, approximations differ from numerical simulations only in small vicinity of  the conjugation point $(x-Vt-r*,V)$, see Figure \ref{final} (right). Anywhere else the approximation suits ideally and is visually undistinguishable from the direct numerical simulations.

Thus the problem of approximation for cylindrical and spherical shock waves is reduced, by the formulas \eqref{appC}, \eqref{appS} and \eqref{r*} to asymptotic  for flat KdV-B shocks. Their analytical presentation is nonexisting, yet they are well-studied. For instance, their oscillatory characteristics can be used.

\section{Stability of constant-boundary cylindrical  and spherical  shocks\label{stable}}

\subsection{Decay of \emph{momentum} conservation law}

As it is well known, for the classical KdV \eqref{KdV} the \emph{momentum} quantity is conserved for solutions rapidly decreasing at infinity:

\begin{equation}\label{mom}
\begin{array}{l}
\frac{\partial}{\partial t} \left.\int_{-\infty}^{+\infty}u^2(x,t)\,dx\right|_{KdV}=
2\int_{-\infty}^{+\infty}u(x,t)\left.\frac{\partial}{\partial t}u(x,t)\right| |_{KdV}\,dx=\\[3mm]
2\int_{-\infty}^{+\infty}u(2uu_x+\lambda u_{xxx})\,dx=
\left.(\frac{4}{3}u^3+\lambda uu_{xx}-  \frac{\lambda}{2}u^2_x\right|^{+\infty}_{-\infty}=0.
\end{array}
\end{equation}

Note that the above argument also shows that  \emph{momentum} is as conserved for zero-boundary conditions solutions on the interval $(+0, +\infty)$: just change $-\infty$ for $+0$ in \eqref{mom}. Moreover, the requirement $u=u_x=0$ on the interval boundaries is sufficient for conservation statement to be true.

But for KdV-B equations \eqref{1} the momentum quantity decay. These equations contain two additional terms with
  respect to  KdV; so  the equation \eqref{mom} now reads

\begin{equation}\label{decay}
\begin{array}{l}
\frac{\partial}{\partial t} \left.\int_{+0}^{+\infty}u^2(x,t)\,dx\right|_{KdV-B}=\\[3mm]
2\int_{+0}^{+\infty}u(x,t)\left.\frac{\partial}{\partial t}u(x,t)\right| |_{KdV-B}\,dx=\\[3mm]
-2\int_{+0}^{+\infty}(\varepsilon u^2_{x}+n t^{-1})u^2)\,dx
-\left.\lambda (u_x)^2\right|_{+0}^{+\infty}.
\end{array}
\end{equation}

The right-hand side of \eqref{decay} is negative.
Thus for the positive
momentum $m(t)=  \int_{+0}^{+\infty}u^2(x,t)\,dx>0$,  its derivative $m_t<0$, so
$m$ decay as long as
\[
\varepsilon u^2_{x}+nt^{-1}u^2+\lambda u_x^2(0,t)\neq 0
\].
It follows that
\[
\lim\limits_{t\rightarrow 0}m(t)=0 \mbox{ and inevitably}
\lim\limits_{t\rightarrow 0}u(x,t)=0
  \]
for any zero-boundary solution of KdV-B on $(0+,+\infty)$.

\subsection{ Stability of KdV-B travelling shocks}

 Let $F_{H}(x,t)$ be a travelling shock wave with $H$ as the boundary condition at $+0$, and a perturbation $v(x)$  of $F_{H}(x,0)$ with the same boundary conditions as $F_H$ (also recall that for shocks $F_H|_{+\infty}=0$).

 Solving
 \eqref{1} for initial value/boundary problem
 \[\{\delta|_{x=+\infty}=0;
\delta|_{x=+0}=0;\delta(x,0)=F_{H}(x,0)-v(x)\}
\]
results in the solution $\delta(x,t)$ with zero boundary conditions:
$\delta=0$ at $x=+0$ and at $x=+\infty$

It was shown above that \emph{momentum} of $\delta$ decays and
\[
\lim\limits_{t\rightarrow+\infty}\delta(x,t)=0.
\]
 It means that the initial difference between $F_{H}(x,0)$ and $v(x)$  vanishes.

 Thus the stability of shocks is proved in the class of rapidly decreasing at infinity cylindrical and spherical KdV-B solutions on the interval $(+0,+\infty)$  with a constant value at $x=+0$.

\begin{figure}[h]
 \begin{minipage}{13pc}
\includegraphics[width=13pc]{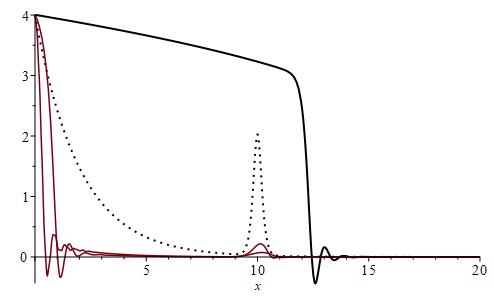}
\end{minipage}
 \begin{minipage}{13pc}
\includegraphics[width=13pc]{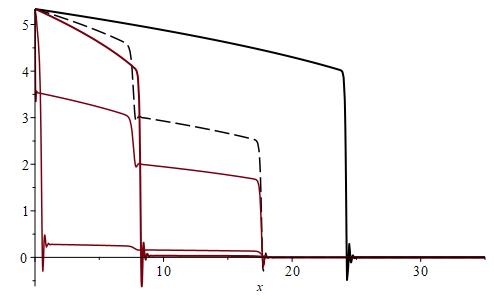}
\end{minipage}
\caption{ \small
\textbf{Left:} Decay of solitary perturbation (dot line); Thick line  --- resulting shock wave ($t=4$), thin lines --- intermediate states
at $t=0.05$ and $t=0.2$.
\textbf{Right:} Evolution of two shocks superposition (dash line); thick lines ---resulting KdV-B shock at $t=2$ and $t=6$;
thin lines --- intermediate states at $t=0.01$ and $t=0.1$.}
\label{stability}
\end{figure}

The Figure \ref{stability} (left) illustrate a typical example; media
$\{\varepsilon=0.25,\lambda=0.05,\}; n=\frac{1}{2}$, $V=3$.

\section{Superposition of shocks  \label{super}}

There are two immediate corollaries from the stability property.

 \subsection{Superposition rule.}

\begin{equation*}
F_{H_1}+F_{H_2}\rightarrow F_{H_1+H_2}
\end{equation*}

  The sum  $F_{H_1}+F_{H_2}$ is not, of course, a solution to the  nonlinear equation KdV-B \eqref{1}.
Yet this sum
 has the boundary values same as
$F_{H_1+H_2}$.

Now just take $\delta(x,0)=[F_{H_1}+F_{H_2}- F_{H_1+H_2}]|_{t=0}$
in the above argument on the
\emph{momentum} decay: $\lim\limits_{t\rightarrow+\infty}\delta(x,t)=0$.

Note that the superposition of two monotonic shocks may be oscillatory but not the other way round. See  Figure \ref{stability} (right) illustrating the evolution of sum two cylindrical shocks in the case of
$\{\varepsilon=0.1,\lambda=0.0075\}$, $V_1=1.5,\, V_2=2.5$.

  \subsection{Superposition Velocity.}
 Nonlinear superposition $F_(H_1)$ and $F_(H_2)$  has the velocity $V_1+V_2$

   Indeed $V$ is proportional to $H$: recall that
   \[H=\frac{4}{3}V, \mbox{ in cylindrical case and } H=\sqrt{e}V
   \mbox { in spherical}
   \]
Thus $H=H_1+H_2$ implies $ V=V_1+V_2$.

\section{Discussion\label{discus}}
 The cylindrical and spherical KdV-B
  differs from the flat KdV-B by an additional  term $n\cdot u/t$.

  As a result, instead of the flat KdV-B travelling shock waves appears a class shock waves with fixed values at $x=+0$ and  rapidly decrease
at $+\infty$.

In this paper a high quality approximations for a one-parameter family of the cylindrical and spherical KdV-B shock waves  is obtained  and their stability is
  proved using the KdV \emph{momentum } conservation law.  Based on these results, the rules of superposition are obtained.

  This study is yet another example of the affectivity of the selective decay approach    to study of diffusion/dissipanive and dispersion effects.

This results are true for diverging waves. Converging waves have their peculiarities and are currently under study.`

  Main ideas and results are illustrated by numerically worked out  graphs;  Maple-21 standard packages: DEtools, PDEtools and Plots were used in this study.

\section*{Funding}
This research received no external funding.

\end{document}